\documentclass[a4paper,11 pt,english]{article}
\usepackage[T1]{fontenc}
\usepackage{times}
\usepackage{multibib}
\newcites{m}{References}

\usepackage{amsmath,amssymb,amsfonts}
\usepackage{graphicx}
\usepackage{caption, subcaption}
\usepackage{abstract}

\usepackage{physics}
\usepackage{upquote}
\usepackage[margin=2.5cm]{geometry}
\usepackage[onehalfspacing]{setspace}
\usepackage{cite}
\usepackage{hyperref}
\usepackage[affil-it]{authblk}
\usepackage[dvipsnames]{xcolor}
\usepackage{sectsty}
\title{Fundamental exciton linewidth broadening in monolayer transition metal dichalcogenides}
\author{Garima Gupta, Kausik Majumdar$^*$\\
Department of Electrical Communication Engineering, Indian Institute of Science, Bangalore 560012, India \\
$^*$Corresponding author, E-mail: kausikm@iisc.ac.in}

\date{}

\begin{document}
\maketitle
\begin{abstract}
	Monolayer Transition Metal Dichalcogenides (TMDS) are highly luminescent materials despite being sub-nanometer thick. This is due to the ultra-short ($<1$ ps) radiative lifetime of the strongly bound bright excitons hosted by these materials. The intrinsically short radiative lifetime results in a large broadening in the exciton band with a magnitude that is about two orders greater than the spread of the light cone itself. The situation calls for a need to revisit the conventional light cone picture. We present a modified light cone concept which places the light line $(\hbar cQ)$ as the generalized lower bound for allowed radiative recombination.  A self-consistent methodology, which becomes crucial upon inclusion of large radiative broadening in the exciton band, is proposed to segregate the radiative and the non-radiative components of the homogeneous exciton linewidth. We estimate a fundamental radiative linewidth of $1.54\pm0.17\ $meV, owing purely to finite radiative lifetime in the absence of non-radiative dephasing processes.  As a direct consequence of the large radiative limit, we find a surprisingly large ($\sim 0.27 $ meV) linewidth broadening due to zero-point energy of acoustic phonons. This obscures the precise experimental determination of the intrinsic radiative linewidth and sets a fundamental limit on the non-radiative linewidth broadening at $T = 0$ K.
\end{abstract}
\pagebreak
\renewcommand{\thesection}{\arabic{section}.}
\renewcommand{\thesubsection}{\thesection.\arabic{subsection}.}
\sectionfont{\color{black}}
\subsectionfont{\color{black}}

\section{INTRODUCTION}
Monolayers of transition metal dichalcogenides (TMDs, for example, MoS\textsubscript{2}, MoSe\textsubscript{2}, WS\textsubscript{2} and WSe\textsubscript{2}) host a unique class of strongly bound two-dimensional excitons \citem{ye2014probing,he2014tightly,ross2013electrical} with a large binding energy of about 500 meV \citem{berkelbach2013theory,chernikov2014exciton,hill2015observation,ye2014probing,gupta2017direct}. This has attracted significant interest from the researchers, and a wide variety of excitonic complexes and their manipulations have been reported in the recent past \citem{hao2017neutral,liu2015strong,amo2010exciton}. Interestingly, these excitons exhibit ultra-short radiative lifetime ($\tau_{r} <1$ ps) \citem{chow2017phonon,engel2018room,robert2016exciton,wang2016radiative}, which is orders of magnitude shorter than that of conventional semiconducting light emitters including III-V semiconductors like GaAs, InGaAs  \citem{sun1996radiative,badcock2016radiative,goswami2012optical}, II-VI semiconductors like CdSe and their quantum dots \citem{goswami2012optical,mohamed2014time,califano2004direct}, organic semiconductors \citem{so1991evidence} and carbon nanotube \citem{perebeinos2005radiative,ouyang2007theoretical,spataru2005theory}. Such a fast radiative-decay manifests as strong photoluminescence \citem{splendiani2010emerging,steinhoff2015efficient} and electroluminescence \citem{cheng2014electroluminescence,sundaram2013electroluminescence} exhibited by the monolayers of these materials. This makes these ultra-thin, flexible, photo-active crystalline sheets an excellent candidate for a plethora of light emitting applications. A deep insight into the fundamental limits of the radiative excitonic decay in these materials is thus essential to exploit the full potential of the short-lived excitons.	

\par One inevitable consequence of the ultra-short radiative lifetime is the large homogeneous broadening (of about 1-2 meV) of the excitonic states due to Heisenberg's uncertainty principle. This broadening is more than two orders of magnitude larger than the total energy extent of the conventional light cone in the exciton band structure. This suggests a need to revisit our understanding of light cone in such systems. In this work, we propose a generalization of the light cone picture to address two fundamental homogeneous broadening mechanisms that determine the lowest achievable excitonic emission linewidth in an experiment, namely: (1) radiative broadening due to ultrashort spontaneous emission lifetime and (2) non-radiative broadening due to zero-point energy of phonons. To address these limiting scenarios, we use a combination of theory and temperature dependent photoluminescence data from high quality monolayer MoSe\textsubscript{2} samples. The limits, as imposed by the aforementioned governing mechanisms on the homogeneous linewidth are estimated to be $\Gamma_{hom,R}=1.54\pm0.17\ $meV and $\Gamma_{NR}(T=0)\approx 0.27\: $meV respectively. This is obtained using a self-consistent approach, that correlates the microscopic exciton band broadening with the experimental data.

\section{REVISITING THE CONCEPT OF LIGHT CONE FOR EXCITONS IN MONOLAYER TMDS}	
\par  \textcolor{black}{The radiative linewidth of 1s exciton photoluminescence peak in monolayer TMDs has been reported to be on the order of 1-2 meV \citem{ajayi2017approaching,cadiz2017excitonic,shepard2017trion} - a manifestation of fast spontaneous radiative recombination.} It is instructive to include this radiative lifetime induced large fundamental broadening of the excitonic states and also contributions from other dephasing mechanisms within the existing light cone understanding. The idea of the conventional light cone in the case of infinite exciton lifetime is the following: \textcolor{black}{For the monolayers of Mo based materials, we only consider contribution from the spin allowed bright transition between the topmost valence band $v$ and the lowermost conduction band $c$ and the  around the $\textbf{K},\textbf{K}^{\prime}$ points in the Brillouin zone (see Figure 1a). We solve the two-particle exciton Hamiltonian in the Bethe-Salpeter (BS) formalism to obtain the exciton band dispersion with the centre of mass (COM) momentum $\textbf{Q} = \textbf{k}_\textbf{e}+\textbf{k}_\textbf{h}$ \citem{wu2015exciton,qiu2015nonanalyticity}. The state $\ket*{{\Psi_{s}(\textbf{Q})}},$ which denotes an exciton state occupying band $s$ at a characteristic momentum $\textbf{Q},$ can be expanded in the basis of single electron and hole states in the reciprocal space as}

\begin{equation}
\ket*{\Psi_{s}(\textbf{Q})} = \sum_{v,c,\textbf{k}} \lambda^{(s)}_{\textbf{Q}}(\textbf{k})\ket*{v,\textbf{k}}\ket*{c,\textbf{k}+\textbf{Q}}
\end{equation}

 The lowest energy $A_{1s}^{0}$ exciton band is schematically illustrated in Figure 1b. On zooming we get Figure 1c, showing that only a small fraction of the states in the band are radiatively bright; the demarcation between the allowed and prohibited states for radiative recombination is determined by the light cone boundary $\textbf{Q}_\textbf{0}$ $(\textbf{Q}_\textbf{0} : E_{1s}(\textbf{Q}_\textbf{0}) = \hbar cQ_0)$. Owing to the energy and in-plane momentum conservation between the recombining exciton and the emitted photon, the states lying at $\textbf{Q}>\textbf{Q}_\textbf{0}$ are radiatively dark, as the out-of-plane component of the momentum of the emitted photons by these excitonic states ceases to be a real value.

\par The energy extent of the light cone is roughly given by $ \Delta E = \hbar^2Q_0^2/2m_{ex} \sim 4$ $\mu$eV (Figure 1c) ($\hbar$ is the reduced Planck constant and $m_{ex}$ is the exciton mass), which is almost two orders of magnitude smaller than the \textcolor{black}{exciton linewidth closest to the homogeneous limit as reported by Cadiz et al. \citem{cadiz2017excitonic} and Ajayi et al. \citem{ajayi2017approaching}}. This mismatch is schematically illustrated in Figure 1d, where the large broadening of the discrete exciton states in the dispersion curve is explicitly shown. This is in stark contrast with Figure 1c which is valid only when the exciton lifetime is very large. The light cone determined radiative boundary now undergoes a modification due to exciton state broadening, extending beyond $\textbf{Q}_\textbf{0}$. An exciton of energy $E_{ex}(\textbf{Q})$ emits a photon with out-of-plane component of momentum $q_z = \sqrt{\left(E_{ex}(\textbf{Q})/\hbar c\right)^{2}-Q^2},$ which is a real quantity provided $E_{ex}(\textbf{Q}) \geq \hbar cQ.$ Hence, the photon dispersion due to its in-plane momentum $\hbar cQ$ (light line) is the lower bound energy for allowed radiative transition of excitons at $\textbf{Q}$ . Figure 1d is a general light cone diagram in TMDs, in which a non-zero fraction of states at $\textbf{Q} > \textbf{Q}_\textbf{0}$ are radiatively active depending on the extent of the broadening, which are otherwise assumed to be forbidden for the conventional light cone in Figure 1c.

\section{EXPERIMENT}
\par We mechanically exfoliate monolayers of MoSe\textsubscript{2} on a clean Si substrate covered with 285 nm thick SiO\textsubscript{2}. \textcolor{black}{Photoluminescence measurement is carried out by varying the sample temperature from 3.2 K to 220 K. The pressure of the sample chamber is kept below $10^{-4}$ torr at all measurement temperatures. A temperature dependent plot of the acquired spectra  upon illumination with a cw laser of wavelength 633 nm is shown in Figure 2a. The optical power density on the sample is kept below $50\ \mu \text{W}/\mu \text{m}^2$ to avoid any laser induced heating effect.} The two conspicuous peaks correspond to the neutral exciton $A_{1s}^{0}$ and the charged trion $A_{1s}^{T}$. The $A_{1s}^{0}$ peak is fitted with a Voigt function and its constituent Lorentzian component obtained upon deconvolution is the homogeneous part of the total broadening \textcolor{black}{\citem{cadiz2017excitonic,engel2018room}}. An example is shown in Figure 2b where the experimental data (in symbols) is the  acquired spectrum at 3.2 K. The red (orange) line shows the Voigt (Lorentzian) fitting with an FWHM of $5.47 (1.92)$ meV. The variation in the homogeneous $(2\Gamma_{hom})$, inhomogeneous $(2\Gamma_{inhom})$ and the total linewidth of the $A_{1s}^{0}$ exciton with temperature is shown in Figure 2c. \textcolor{black}{The analysis in the following sections is derived from the extracted homogeneous component of the excitonic linewidth only.}

\par When compared with previous experimental data \citem{ajayi2017approaching,cadiz2017excitonic} and theoretical calculations \citem{singh2015intrinsic}, it is evident that our lowest obtained $1.92$ meV homogeneous linewidth is predominantly due to radiative population decay of the excitons. We further affirm this assertion by calculating the photoluminescence spectrum for the limiting case of broadening purely because of fundamental radiative lifetime (as explained later),  which has a spectral linewidth $(2\Gamma_{hom,R})$ of 1.54 meV, as shown in Figure 2b (in green).

\section{SELF-CONSISTENT METHODOLGY TO SEGREGATE RADIATIVE AND NON-RADIATIVE LINEWIDTH COMPONENTS}	
\par \textcolor{black}{The aim of this section is to segregate the radiative and non-radiative components of the homogeneous exciton linewidth, given an experimentally obtained photoluminescence spectrum.} Previous works have reported calculation of radiative lifetime of $A_{1s}^{0}$ exciton assuming zero \citem{palummo2015exciton, chen2018theory} and finite \citem{wang2016radiative} energy broadening in the exciton band. We shall prove in the following text that in order to obtain the fundamental limits in question, it is important to take into account both the radiative and the non-radiative broadening mechanisms. To achieve this, we propose a self-consistent framework for the calculation of the radiative lifetime of excitons in TMDs. We take $\Gamma(\textbf{Q})=\Gamma_{R}(\textbf{Q})+\Gamma_{NR},$ where $\Gamma_{R}(\textbf{Q})$ and $\Gamma_{NR}$ are induced by radiative population decay and non-radiative dephasing respectively at a given $\textbf{Q}.$ Note that we assume $\Gamma_{NR}$ to be independent of $\textbf{Q}$.  $\Gamma_{R}(\textbf{Q})$ (and hence the exciton radiative lifetime $\tau_{R}(\textbf{Q})$ ) is then given by:

\begin{equation}
\begin{split}
\Gamma_{R}(\textbf{Q}) &= \dfrac{\hbar}{2\tau_{R}(\textbf{Q})} \\
&=\eta_{0}\dfrac{\hbar e^{2}}{2m_{0}^{2}}|\chi_{ex}(\textbf{Q})|^{2}\int_{0}^{\infty}dq_{z}\dfrac{1}{\sqrt{Q^{2}+q_{z}^{2}}}\times\left(1+\dfrac{q_{z}^{2}}{q_{z}^{2}+Q^{2}}\right)\\
&\times\dfrac{(\Gamma_{R}(\textbf{Q})+\Gamma_{NR})/\pi}{\left[E_{ex}(\textbf{Q})-\hbar c \sqrt{Q^{2}+q_{z}^{2}}\right]^{2}+(\Gamma_{R}(\textbf{Q})+\Gamma_{NR})^{2}}\\
%& = \dfrac{2\Gamma_{RAD}(\vec{Q})}{\hbar}\\
%& = \dfrac{1}{\hbar/(2\Gamma_{RAD}(\vec{Q}))}
\end{split}
\end{equation}

Here $\eta_{0}$ is the free space impedance and the function $\chi_{ex}(\textbf{Q})$ is obtained by integrating the quantity $\textbf{P}_{vc,\textbf{Q}}(\textbf{k}).\lambda^{(s)}_{\textbf{Q}}(\textbf{k})$ in the $\textbf{k}$ space \citem{zhang2014absorption}, where $\textbf{P}_{vc,\textbf{Q}}(\textbf{k})$ is the momentum matrix element between $\ket*{v,\textbf{k}}$ and $\ket*{c,\textbf{k}+\textbf{Q}}$ \citem{xiao2012coupled}. The quasiparticle bandstructure for monolayer MoSe\textsubscript{2} is obtained using the Lowdin Hamiltonian \citem{kormanyos2015k, kormanyos2013monolayer}. Details of the Hamiltonian, BS equation and calculation of $|\chi_{ex}(\textbf{Q})|^2$ are provided in the Supporting Note 1. The right panel of Figure 1d is a cartoon plot of the decay rate of excitons with energy at different  $\textbf{Q}$ points. We define $I_{cal}^{1s}$ as the calculated homogeneous output of the spontaneous emission from the 1s excitons as would be obtained in a PL experiment:

\begin{equation}
I_{cal}^{1s}(E) \propto \sum_{\textbf{Q}}\dfrac{1}{(\exp^{E_{1s}(\textbf{Q})/k_{B}T}-1)}\cdot\dfrac{1}{\tau_{R}(E,\textbf{Q})}\;\;(\text{where}\; E \geq \hbar cQ)
\end{equation}

\par We work in the mathematical framework of equations (2) and (3) put together ensuring that the calculated broadening of $I_{cal}^{1s}$ matches the experimental Lorentzian linewidth of the $A_{1s}^{0}$ exciton. \par The flow chart in \textcolor{black}{Appendix A (Figure 5)} explains our proposed methodology for the quantitative estimation of $\Gamma_{R}(\textbf{Q})$ and $\Gamma_{NR}\;$ from $\Gamma_{hom,exp}$ (Figure 2c). It works in the following way: on starting with an assumed value $\Gamma_{NR}^{(0)},$ the first step is the radiative lifetime calculation. Note that the transcendental nature of equation (2) requires solving $\Gamma_{R}^{(i)}(\textbf{Q})$ self-consistently for a given $\Gamma_{NR}^{(i)}$, as shown in the right block, where $i$ stands for the iteration index. We next obtain $I_{cal}^{1s}$ (equation 3) with $\tau_{R}^{(i)}$ calculated for the converged values of $\Gamma_{R}^{(i)}(\textbf{Q}).$  $\Gamma_{NR}^{(i)}$ is updated appropriately for the successive iterations on comparing the calculated $I_{cal}^{1s}$ linewidth $(2\Gamma_{hom,cal})$ with $2\Gamma_{hom,exp}$. In summary, the left block ensures matching of experimental data with the calculated linewidth and the right block takes into account the necessary self-consistency of equation (2) at a given value of $\Gamma_{NR}$.

\section{DISCUSSIONS}
\renewcommand{\thesection}{\arabic{section}}
\subsection{Fundamental Limit of Radiative Linewidth}
\par The self consistent nature of equation (2) becomes more important when $\Gamma_{R}(\textbf{Q})$ is comparable or larger than $\Gamma_{NR}.$ This is precisely the case for monolayer TMDs at low temperature and at low excitation density. In the limiting case of $\Gamma_{NR} \ll \Gamma_{R}(\textbf{Q}),$ we can solve for $\Gamma_{R}(\textbf{Q})$ independently from equation (2). This leads to a fundamental homogeneous emission linewidth of $1.54\pm0.17\ $meV as shown by the green line in Figure 2b. This shows that the exciton Hamiltonian is sufficiently non-Hermitian due to the fast exciton radiative decay \citem{molina2016temperature}, thereby validating the need for such self-consistent approach for determining the exciton radiative lifetime even in the absence of any non-radiative scattering process. The error bar takes into account a  $\pm 5\% $ variation in the Lowdin Hamiltonian parameters and sensitivity of $\textbf{k}$ space griding in calculating $|\chi_{ex}(\textbf{Q})|^2$ \citem{qiu2013optical}. Note that the technique is an extremely powerful tool to obtain the fundamental radiative linewidth as it does not require any broadening to be introduced `by hand', as is usually done in most calculations, rather it self-consistently takes care of the radiative component of the homogeneous broadening.

\par The variation in $\Gamma_{hom}$ as a function of the extracted $\Gamma_{NR}$ at each temperature using the proposed algorithm is plotted in Figure 3a. The data points track closely a reference line of slope $\sim 0.98$ as shown in the figure, thus affirming the non-radiative nature of the homogeneous linewidth at all temperatures. The inset plot is a zoomed in view that shows, on extrapolation, a residual radiative linewidth of $1.58$ meV corresponding to $\Gamma_{NR} = 0.$ This is consistent with our calculated fundamental linewidth value of $1.54\pm0.17\ $meV.

\subsection{Temperature Dependence of Radiative Linewidth - Inside and Outside Light Cone}
\par Figure 3b illustrates the temperature induced variation in the self-consistently obtained $\textbf{Q}$ resolved  $\Gamma_{R}$. The corresponding values of the extracted $\Gamma_{NR}$ is shown in the legend.  For reference, the limiting case of purely radiative linewidth at $T = 0$ K and $\Gamma_{NR} = 0$ meV is shown in black. $\Gamma_{R}(\textbf{Q})$ shows a strongly non-monotonic behavior with $\textbf{Q}$, reaching its maximum value at $\textbf{Q}_\textbf{0}^{'}$. The color shades represent the different characteristic regions. $\Gamma_{R}(\textbf{Q})$ is almost invariant for the most part of $\textbf{Q} < \textbf{Q}_\textbf{0}^{'}$ points within the light cone (white region). With an increase in temperature, $\textbf{Q}_\textbf{0}^{'}$ shifts towards $\textbf{Q}=\textbf{0}$. This can be understood as a competition between two temperature dependent effects, namely a reduction in the optical bandgap ($A_{1s}^{0}$) and an increase in $\Gamma_{NR}$ with $T$. This is schematically explained in Figure 3c for three different temperatures with $0 < T_1 < T_2$. For $\textbf{Q}_\textbf{0}^{'}<\textbf{Q}<\textbf{Q}_\textbf{0}^{''}$ (green shade), owing to larger $\Gamma_{NR}$ at higher $T$, a larger fraction of the states lie below the light line, which manifests as a strong suppression in $\Gamma_{R}(\textbf{Q})$. All the curves converge to a point $\textbf{Q}_\textbf{0}^{''}\approx \textbf{Q}_\textbf{0}(0)$, before diverging again in the blue region. Note that, the contribution of the state $\textbf{Q}_\textbf{0}(T)$ remains $50\%$ irrespective of temperature. For $\textbf{Q} >\textbf{Q}_\textbf{0}^{''}$ (blue region),  $\Gamma_{R}(\textbf{Q})$ experiences a significant increment at higher $T$ as more excitonic states appear above the light line $\hbar cQ$, which are otherwise forbidden in the case of negligible broadening. The variation of $\Gamma_{R}(\textbf{Q})$ as a function of temperature is explained in Supporting Note 2.

\subsection{Non-Radiative Linewidth Broadening and its Fundamental Limit}
The extracted $\Gamma_{NR}$ is plotted in Figure 4a (symbols) as a function of temperature. Such a strong temperature dependence is well described by exciton-phonon scattering processes \citem{singh2015intrinsic,selig2016excitonic}. As the radiative emission process is a result of recombination of excitons $\textbf{Q}\approx \textbf{0}$, one usually assumes that only phonon absorption processes are allowed \citem{hellmann1993homogeneous}. However, in the presence of large radiative broadening of the excitonic states, the low energy (close to zone-center) acoustic phonon emission process is allowed within the light cone, apart from the usual absorption processes of acoustic and optical phonons. Therefore, the temperature dependence of the induced non-radiative broadening is given by
\begin{align}
\Gamma^{cal}_{NR} &= \sum_{\Omega_{AP}}C_{AP}\;[2N(\Omega_{AP},T)+1] + \sum_{\Omega_{OP}}C_{OP}\;N(\Omega_{OP},T)\\
&= \underbrace{\sum_{\Omega_{AP}}C_{AP}}_{\Theta_{0}} + \underbrace{\sum_{\Omega_{AP}}2C_{AP}\;N(\Omega_{AP},T)}_{\Theta_{AP}} + \underbrace{\sum_{\Omega_{OP}}C_{OP}\;N(\Omega_{OP},T)}_{\Theta_{OP}}
\end{align}
Here $N$ denotes the Bose occupation number given by $N(\Omega, T) = (e^{\frac{\hbar \Omega}{k_{B}T}}-1)^{-1}$. The combination of the second and the third terms in equation 5 is shown in red line in Figure 4a, showing good agreement with the data at temperature higher than $50$ K. We take $\Omega_{AP}$ values for modes in the acoustic branch up to 100 cm\textsuperscript{-1} with a fitted coupling strength $C_{AP} = 162.5\; \mu $eV. $\;\Omega_{OP}$ corresponds to specific optical modes of frequency $150, 195, 240, 275,$ and $290$ cm\textsuperscript{-1} with coupling strength $C_{OP} = 1, 2.5, 5, 9,$ and $13.75$ meV, respectively. The individual mode contributions to $\Gamma_{NR}$ is also shown in the same plot. The acoustic phonon branch contributes almost linearly to $\Gamma_{NR}$ whereas the optical modes are responsible for the superlinear increase in broadening at higher temperatures.

At lower temperatures ($T<50$ K), we observe a large deviation between the extracted $\Gamma_{NR}$ and the fitted line using $\Theta_{AP}+\Theta_{OP}$ from equation 5. This residual broadening can originate from multiple non-radiative channels, such as excitation induced [for example, exciton-free carrier scattering \citem{honold1989collision,hellmann1993homogeneous} and exciton-exciton scattering \citem{honold1989collision,hellmann1993homogeneous,singh2015intrinsic}]and spin flip induced dephasing. The spin flipping time being on the order of nanosecond \citem{yang2015long} can only provide a linewidth broadening of a few $\mu$eV and can be safely ignored. We also note that our measurements are performed at a low excitation density of $N_{x}< 9\times 10^8$ cm\textsuperscript{-2}\textcolor{black}{(an upper limit on the exciton density assuming $100\%$ quantum yield)}. In addition, excitation induced broadening ($\propto a_B^2E_bN_x$ where $a_B$ is the Bohr radius and $E_b$ is the binding energy) is suppressed in the case of the $A_{1s}^{0}$ exciton for monolayer TMDs due to small $a_B$ (less than a nanometer). We estimate an overall linewidth broadening of only $\sim 2.4\;\mu$eV due to excitation induced broadening. Further, we discard any contribution of exciton-free carrier scattering in the homogeneous exciton linewidth due to the absence of a uniform free carrier density in monolayer MoSe\textsubscript{2} at low sample temperature. Note that the origin of trion peak in Figure 2a is only a result of inhomogeneous local doping effect induced charge puddles due to charge fluctuations in the SiO\textsubscript{2} substrate. This is verified by the complete Gaussian (and hence inhomogeneous) nature of the line shape of the trion peak. Therefore, any scattering with the localized charge carriers only affects the inhomogeneous broadening of the exciton linewidth, leaving the homogeneous linewidth component unaffected.

We understand the observed deviation at low temperature as the contribution of zero point energy of the acoustic phonons, as described by the first term ($\Theta_0$) in equation 5. At low temperature, following the approach of Marini \citem{marini2008ab}, we obtain a quantitative expression for $\Gamma_{NR}$ given by $\Gamma_{NR}(T)=\alpha\left(\dfrac{E_{0}}{2k_{B}}+T\right),$ where $E_{0}$ is roughly equal to the broadening of the exciton band (derivation in \textcolor{black}{Appendix B}). On putting $E_{0} = 2(\Gamma_{hom,R} + \Gamma_{NR}(0))$, we obtain the fundamental non-radiative linewidth at $T = 0$ K by
\begin{align}
\Gamma_{NR}(0) &= \alpha\left(\dfrac{E_{0}}{2k_{B}}\right),\\
&=\alpha\left(\dfrac{\Gamma_{hom,R}+\Gamma_{NR}(0)}{k_{B}}\right)\\
&=\dfrac{\alpha^{'}}{1-\alpha^{'}}\Gamma_{hom,R} \hspace{2cm}(\alpha^{'}=\dfrac{\alpha}{k_{B}})
\end{align}
The low temperature regime is shown separately in Figure 4b. The slope of $\Gamma_{NR}$ versus $T$ gives an estimate of $\alpha \approx 13.56\; \mu$evK\textsuperscript{-1}. Using the relation in equation 8 and putting the value of $\Gamma_{hom,R}$ as $1.54$ meV, this readily allows us to obtain $\Gamma_{NR}(0) \approx 134\: \mu$eV. In the light of the negligible contribution due to spin-flip and excitation induced dephasing, this is in excellent agreement with the intercept of $150\: \mu$eV of the fitted dashed line at $T=0$ in Figure 4b. This allows us to validate the assertion of large linewidth broadening resulting from zero-point energy and results in a fundamental non-radiative exciton linewidth broadening of $\sim 0.27\: $meV.  An increase in the radiative linewidth directly enhances the zero-point broadening, as suggested by the strong correlation between the two quantities in Equation 8.

\renewcommand{\thesection}{\arabic{section}.}
\section{CONCLUSION}
\par In summary, we have investigated the principal factors that determine the fundamental limit of excitonic linewidth in monolayer TMDs - a class of systems that exhibit extremely short radiative spontaneous decay of strongly bound excitons. We have shown that the large broadening of the excitonic states due to the strong radiative dephasing must be incorporated in the light cone picture in a self-consistent way for accurate estimation of the emission linewidth. We have proposed a powerful technique to segregate the individual radiative and non-radiative components of the linewidth. This self-consistent approach sets a limit of $1.54$ meV on the fundamental radiative linewidth. One striking observation of this limit is that the zero-point energy induced broadening, which determines the fundamental limit of the non-radiative broadening is extraordinarily large ($\sim 0.27\: $meV) - a manifestation of the large radiative broadening that allows the acoustic phonon emission process at the ultra-low temperature regime. \textcolor{black}{The results presented in this work are robust against the inhomogeneity of the sample as long as the factors that cause inhomogeneous broadening does not affect the homogeneous linewidth}. The findings advance the microscopic understanding of light emission governed by tightly bound excitons in layered semiconductors and can pave way for novel optoelectronic device concepts, including exploitation of strong coupling regime of light-matter interaction.

\sectionfont{\color{black}}
\subsectionfont{\color{black}}
\section*{SUPPORTING INFORMATION}
%\textbf{Supporting Information}\\
Supporting Information is available on exciton bandstructure calculation in MoSe\textsubscript{2} and effect of temperature on radiative broadening.

\section*{ACKNOWLEDGMENTS}
This work was supported in part by a grant under Indian Space Research Organization (ISRO), by the grants under Ramanujan Fellowship, Early Career Award, and Nano Mission from the Department of Science and Technology (DST), and by a grant from MHRD, MeitY and DST Nano Mission through NNetRA.
\section*{NOTES}
The authors declare no competing financial interest.

\pagebreak
\begin{figure*}[!hbt]
	%	\centering
%	\vspace{-1.5in}
	\hspace{-0.3in}
	\includegraphics[scale=0.52] {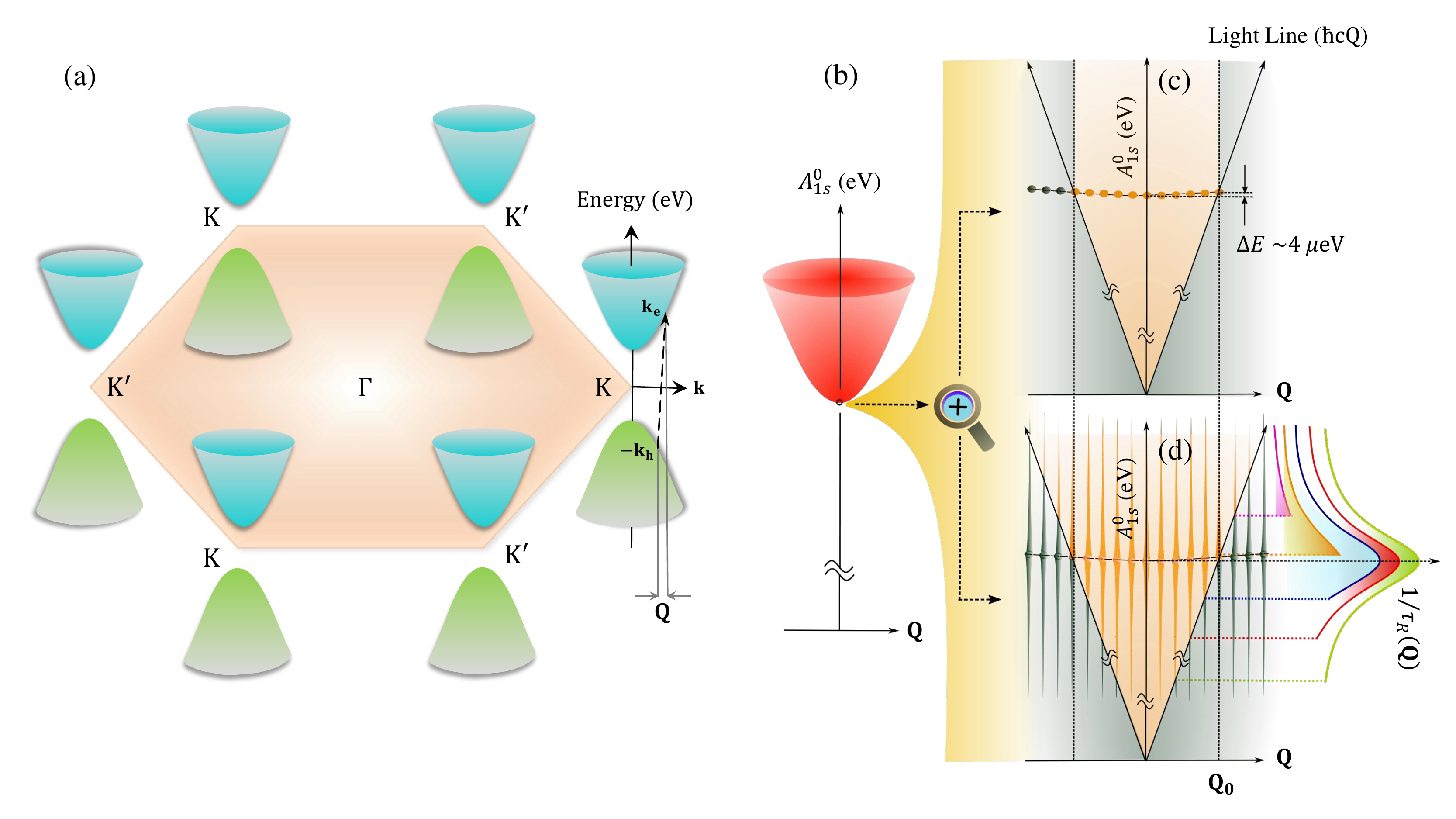}
	% \vspace{-0.1in}
	\caption{\textbf{\textcolor{black}{Modified light cone for exciton bands with large broadening in monolayer TMDs}} \textcolor{black}{(a) Electronic bandstructure of monolayer MoSe\textsubscript{2} around the Brillouin zone corners $\textbf{K},\textbf{K}^{\prime},$ showing the spin allowed bright transition from the uppermost valence band $v$ (in green) to the lowermost consuction band $c$ (in blue) on absorbing a photon. Here $\textbf{k}$ denotes the quasiparticle momentum.} (b) Energy dispersion of the $A_{1s}^{0}$ exciton with its center of mass momentum $(\textbf{Q})$. The conventional light cone for the $A_{1s}^{0}$ exciton shown as zoomed in (c). (c) The radiatively bright exciton states within the light cone are shown as orange dots, while the dark states outside the light cone are represented by grey dots. The demarcation between the two is the light cone boundary $\textbf{Q}_{\textbf{0}}$. The extent of the light cone is $\Delta E=E_{1s} (\textbf{Q}_{\textbf{0}} )- E_{1s} (\textbf{0})  \sim 4 \mu$eV, which is approximately two orders of magnitude smaller than the \textcolor{black}{intrinsic radiative linewidth limit of excitons, which is typically on the order of   1-2 meV}.  (d) The modified light cone showing the large broadening of the excitonic states. The photon energy $\hbar c\text{Q}$ \textcolor{black}{(Light Line)} due to its in-plane component of momentum is the lower bound on the exciton energy for light emission, i.e. excitons with energy $E_{1s} (\textbf{Q})\geq \hbar c\text{Q}$ will emit a photon upon recombination. This extends the conventional light cone beyond $\textbf{Q}_{\textbf{0}}$. (Right panel)$\-$ Decay rate variation of excitons with energy at different $\textbf{Q}$.}\label{fig:F1}
	\end{figure*}

\pagebreak
\begin{figure*}[!hbt]
	%	\centering
	%	\vspace{-1.5in}
	\hspace{0.6in}
	\includegraphics[scale=0.5625] {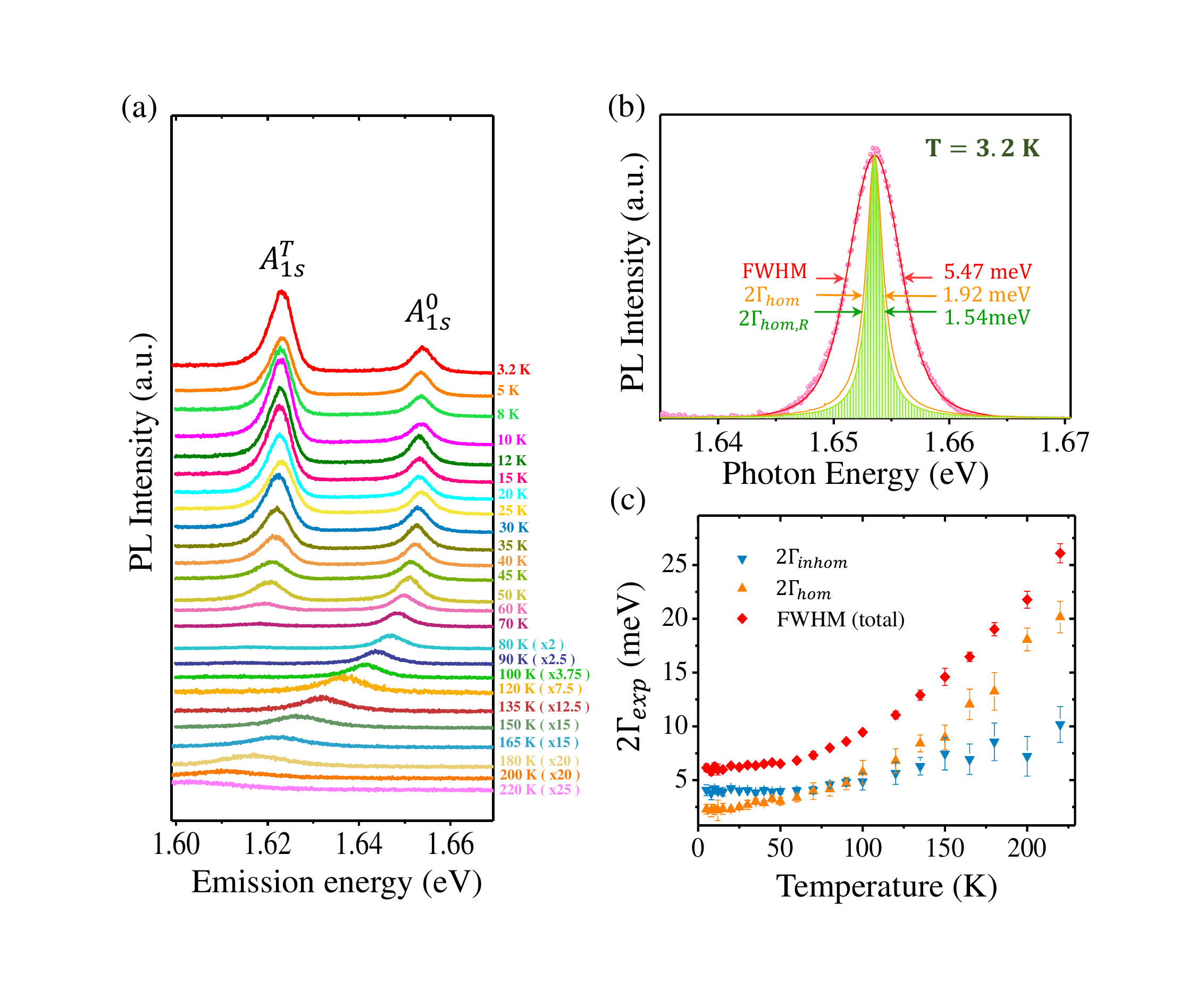}
	% \vspace{-0.1in}
	\caption{\textbf{\textcolor{black}{Photoluminescence linewidth of monolayer MoSe\textsubscript{2}.}} (a) Acquired photoluminescence spectra of monolayer MoSe\textsubscript{2} on SiO\textsubscript{2} substrate showing the 1s neutral exciton $(A_{1s}^{0})$ and trion $(A_{1s}^{T})$ peaks as the sample temperature varies from $3.2$ K to $220$ K. (b) The experimental $A_{1s}^{0}$ spectrum (in symbols) with an $\text{FWHM}=5.47$ meV, the fitted Voigt function (in red line), and the deconvoluted Lorentzian with linewidth of  $2\Gamma_{hom}=1.92$ meV (in orange) at $T = 3.2$ K. The patterned filled curve (in green) is the simulated PL spectrum with broadening equal to the fundamental radiative linewidth of  $2\Gamma_{hom,R}=1.54\;m$eV. (c) Temperature dependence of the extracted $A_{1s}^{0}$  exciton linewidth (FWHM) and its constituent Gaussian (inhomogeneous) $2\Gamma_{inhom}$ and Lorentzian (homogeneous) $2\Gamma_{hom}$  components.}\label{fig:F2}
\end{figure*}
\pagebreak
\begin{figure*}[!hbt]
	%\centering
	\vspace{-2cm}
	\hspace{-0.75in}
	\includegraphics[scale=0.6] {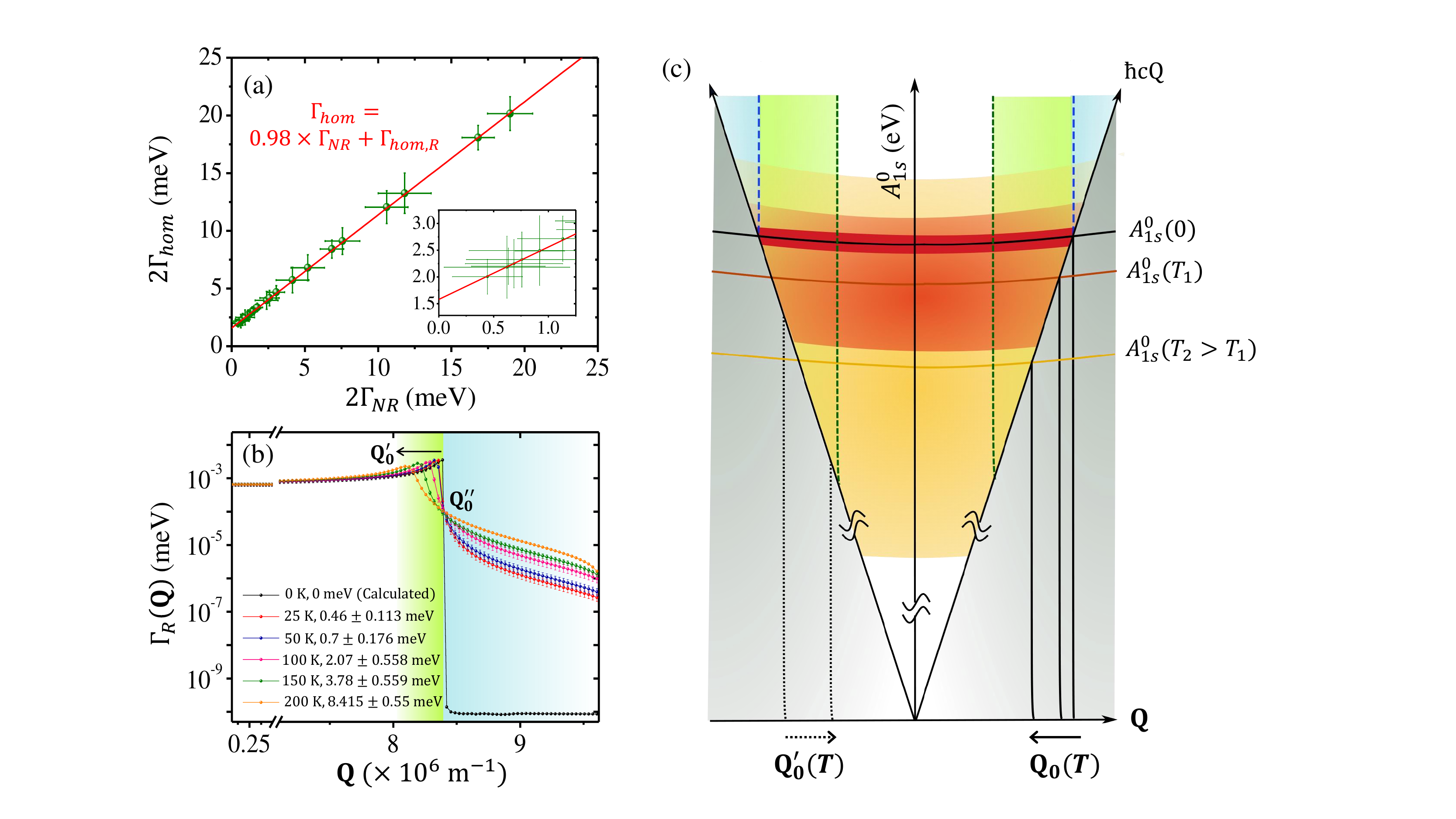}
	%\vspace{-0.1in}
	\caption{\textbf{Temperature dependence of radiative broadening}.(a) The homogeneous exciton linewidth $2\Gamma_{hom,exp}$ variation with pure non-radiative linewidth $2\Gamma_{NR}$ \textcolor{black}{ (extracted from the experimental data)}.  The (red) line with slope close to unity $(\sim 0.98)$ is a good fit to the experimental data. As shown in the inset, an extrapolated value of $1.58$ meV at $\Gamma_{NR}=0$ is consistent with the $2\Gamma_{hom,R}= 1.54\pm0.17$ meV.  (b) Extracted value of $\Gamma_{R}(\textbf{Q})$ at different sample temperatures. The value of $\Gamma_{NR}$ at the corresponding temperature is mentioned in the legend. $\Gamma_{R}(\textbf{Q})$ is maximum at $\textbf{Q}_\textbf{0}^{'}$ and the curves converge at $\textbf{Q}_\textbf{0}^{''}$ irrespective of $T$. (c) Schematic representation of the light cone for three different temperatures $0 < T_1 < T_2$ with the increasing broadening indicated by red, orange and yellow shades. With an increase in $T$, optical bandgap reduces and both $\textbf{Q}_\textbf{0}$ and $\textbf{Q}_\textbf{0}^{'}$ shift towards $\textbf{Q}=\textbf{0}$. The color coding of different regions in $\textbf{Q}$  correspond to (b), namely, white $(\textbf{Q} < \textbf{Q}_\textbf{0}^{'})\-$, green $(\textbf{Q}_\textbf{0}^{'}<\textbf{Q}<\textbf{Q}_\textbf{0}^{''})$ and blue $(\textbf{Q}>\textbf{Q}_\textbf{0}^{''})$.}\label{fig:F3}
	
\end{figure*}

\pagebreak

\begin{figure*}[!hbt]
	\centering
	\vspace{-0.75cm}
	%\hspace{-2.075cm}
	\includegraphics[scale=0.5] {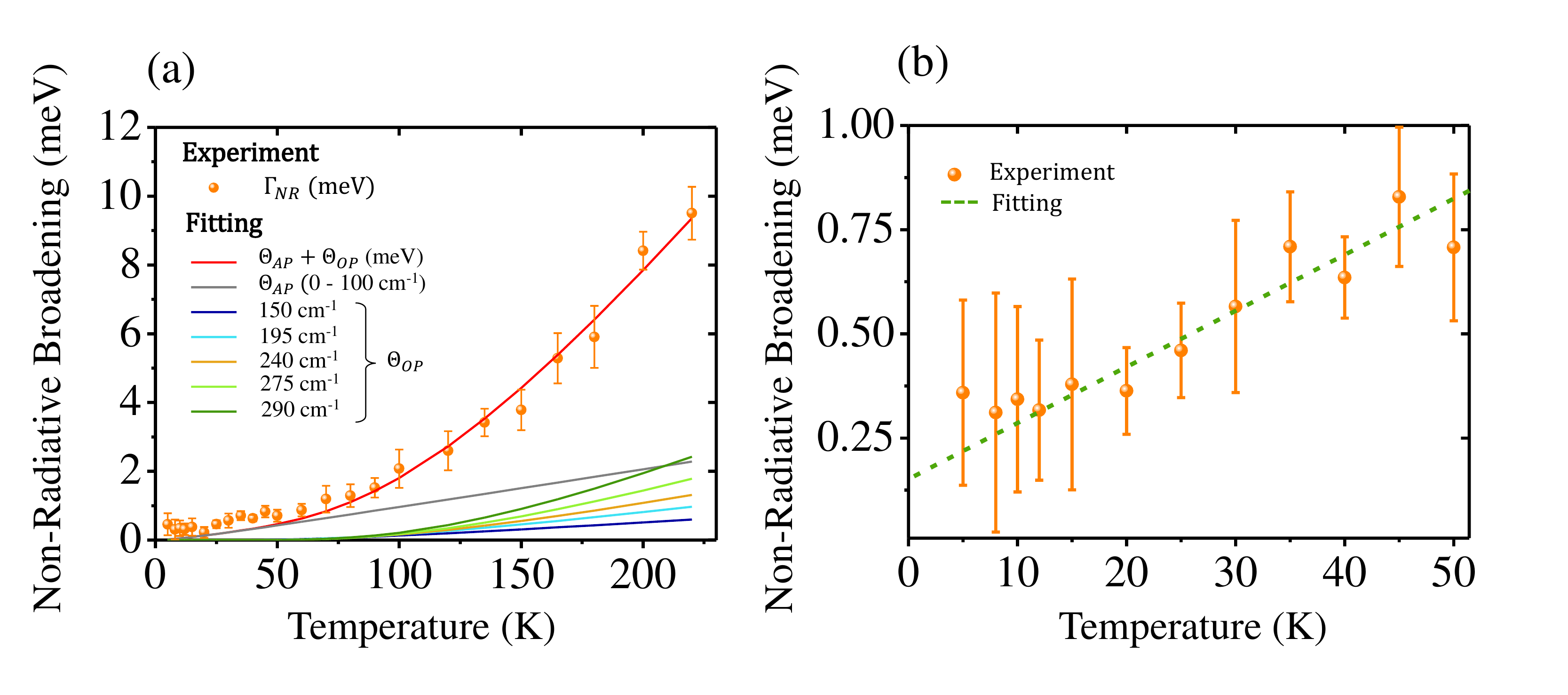}
	%\vspace{-0.1in}
	\caption{\textbf{Non-radiative broadening due to exciton-phonon coupling and zero-point energy.} (a) Extracted $\Gamma_{NR}$ (orange symbols with error bars) and the fitted curve (red line) using a combination of $\Theta_{AP}$ (acoustic phonon emission and absorption) and $\Theta_{OP}$ (optical phonon absorption) terms from equation 4. The individual contribution of acoustic and optical phonons is also shown separately. (b) Zoomed-in view of the low temperature regime [$T<50$  K]. The green dashed line shows a linear fit with a slope of $13.56\; \mu$evK\textsuperscript{-1} and a vertical intercept at $T=0$ K being $\Gamma_{NR}(0) = 150\;\mu$eV. This indicates a large fundamental non-radiative linewidth broadening due to zero-point energy of acoustic phonons.}\label{fig:F4}
\end{figure*}
\pagebreak
\appendix

\sectionfont{\color{black}}
\section*{Appendix A}
\markboth{\MakeUppercase{Appendix}}{\MakeUppercase{Appendix}}
\renewcommand*{\thesection}{A.\arabic{section}}
\appendix
%\renewcommand{\thefigure}{A\arabic{figure}}
%\setcounter{figure}{0}
%\subsection{Flow chart of the algorithm for calculation of radiative and non-radiative components of homogeneous linewidth}
The algorithm employed for calculating exciton radiative lifetime $(\tau_R(\textbf{Q}))$  by deconvoluting the $\textbf{Q}$ resolved radiative $\Gamma_R(\textbf{Q})$ and non-radiative $\Gamma_{NR}$ components, given a homogeneous linewidth obtained from experiment $(2\Gamma_{hom,exp})$. The convergence of the right block ensures a self-consistent $\Gamma_{R}(\textbf{Q})$ for a given $\Gamma_{NR}$ (equation 2). The left block finds the fitted value of $\Gamma_{NR}$ for a given experimentally obtained spectral linewidth of the exciton line.  $\Gamma_{NR}$ is the only fitting parameter in the algorithm.%\label{fig:A1}
\begin{figure*}[!hbt]
%\centering
%	\vspace{-3 cm}
\hspace{-1in}
\includegraphics[scale=0.6] {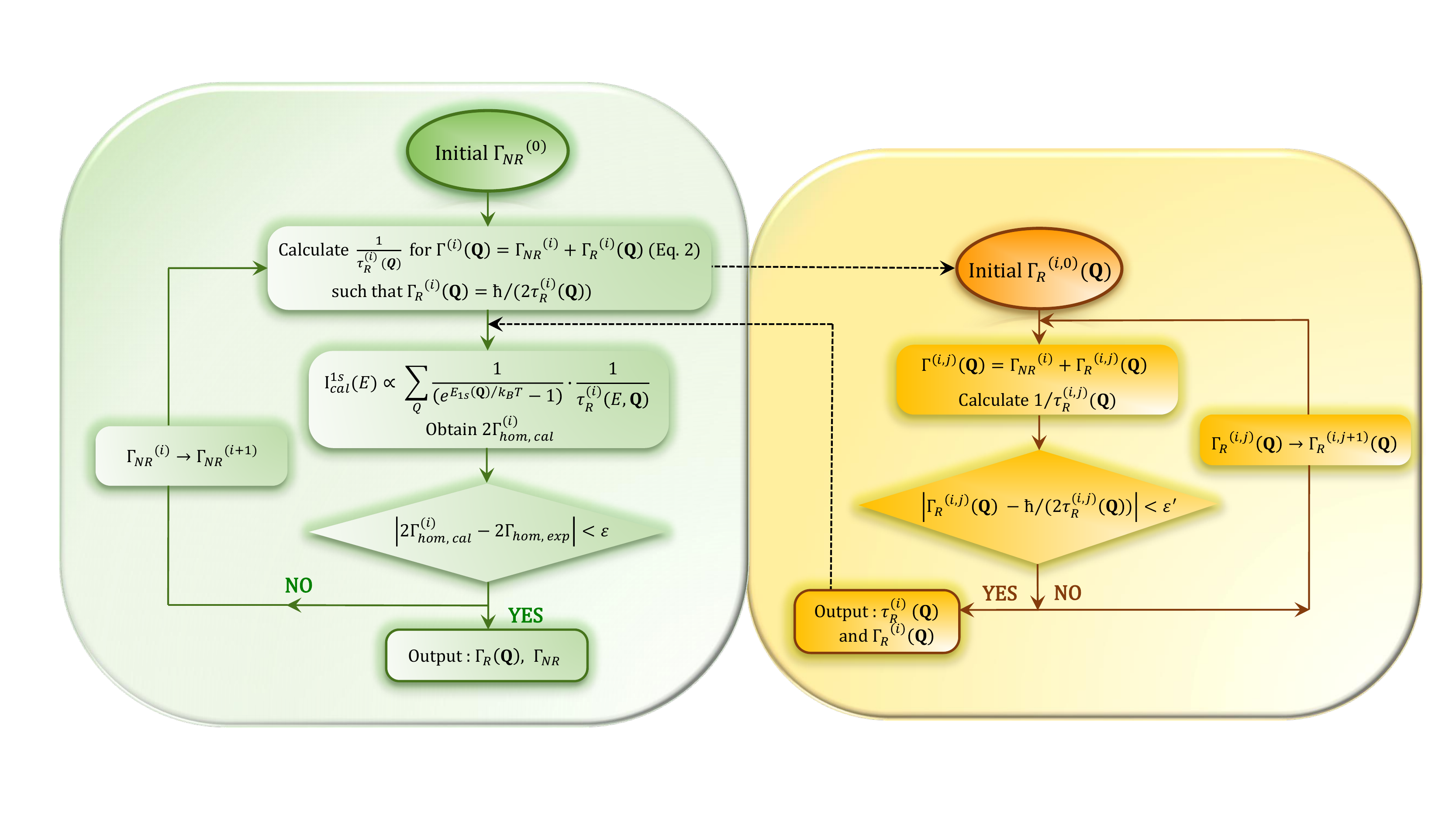}
%\vspace{-0.1in}
\caption{\textbf{Flow chart of the algorithm for calculation of radiative and non-radiative components of homogeneous linewidth}}%\label{fig:time}
\end{figure*}
	\pagebreak
\section*{Appendix B}
The temperature induced broadening of the exciton bands arising due to exciton-phonon scattering \citem{marini2008ab} is given by
\begin{align}
\Gamma_{NR}\:(T) &= \textrm{Im}[E_s(T)]\\
&= \int d\omega\; \textrm{Im}[g^2F_{s}(\omega , T)][N(\omega , T)+\frac{1}{2} \pm \frac{1}{2}]
\end{align}
The upper and the lower signs correspond to phonon emission and absorption processes respectively. $g^2F_{s}(\omega , T) = \sum_{v,c,\textbf{k}} |\lambda^{(s)}(\textbf{k},T)|^2[g^2F_{c,\textbf{k}}(\omega)-g^2F_{v,\textbf{k}}(\omega)]$ is the exciton-phonon coupling function. We consider contribution from only the lowermost conduction band and
the topmost valence band for monolayer MoSe\textsubscript{2}.

\begin{align}
g^2F_{c,\textbf{k}}(\omega)-g^2F_{v,\textbf{k}}(\omega) &= \sum\left(\frac{\partial \epsilon_{c,\textbf{k}}}{\partial N(\omega_{\nu},T)}-\frac{\partial \epsilon_{v,\textbf{k}}}{\partial N(\omega_{\nu},T)}\right)\delta (\omega - \omega_{\nu})\\
&\approx \sum\frac{\partial \epsilon_{g,\textbf{k}}}{\partial N(\omega_{\nu},T)}\delta (\omega - \omega_{\nu})
\end{align}
where the summation runs for all the phonon modes and $\epsilon_{g,\textbf{k}}= \epsilon_{c,\textbf{k}}- \epsilon_{v,\textbf{k}}.$
\begin{align}
\Gamma_{NR}\:(T) = \sum_{\nu,\omega_{\nu}}\textrm{Im}\left[\sum _{\textbf{k}}|\lambda^{(s)}(\textbf{k},T)|^2\frac{\partial \epsilon_{g,\textbf{k}}}{\partial N(\omega_{\nu},T)}\right](N(\omega_{\nu},T)+\frac{1}{2} \pm \frac{1}{2})
\end{align}
We use the shorthand notation $\beta_{\omega_{\nu}}(T) = \textrm{Im}\left[\sum _{\textbf{k}}|\lambda^{(s)}(\textbf{k},T)|^2\dfrac{\partial \epsilon_{g,\textbf{k}}}{\partial N(\omega_{\nu},T)}\right],$  quantifying the coupling strength of excitons to phonon mode $\nu, \omega_{\nu}$ at $T.$ $\:\lambda^{s}(\textbf{k},T)$ is the BSE solution of the non-Hermition exciton Hamiltonian, the temperature dependence arising as a result of quasiparticle bands broadening due to electron phonon scattering. The exciton phonon scattering does not vanish even at $T = 0$ K, as $N \rightarrow 0,$ due to scattering via phonon emission process and induces finite zero point broadening in the exciton band.
\par In order to quantify $\Gamma_{NR}$ due to zone center acoustic phonon emission at low temperatures, we approximate for analytical purpose, that the coupling coefficient $\beta_{\omega_{\nu}}(T)$ is invariant with $\omega_{\nu}, T$ and replace $\beta_{\omega_{\nu}}(T)$ by $\beta$. On putting  $N\approx\dfrac{k_{B}T}{\hbar\omega_{\nu}}$ for the acoustic phonon modes at small temperatures, we get

\begin{align*}
\Gamma_{NR}\:(T)
& = \beta\sum_{\omega_{\nu}}\left(\dfrac{k_{B}T}{\hbar\omega_{\nu}}+1\right)\\
& = \dfrac{\beta A N_{b}}{2\pi}\int_{0}^{q_{0}}dq\: q\left(\dfrac{k_{B}T}{\hbar cq}+1\right) \hspace{1cm}( \hbar \omega_{\nu} = \hbar cq)\\
& = \dfrac{\beta AN_{b}}{2\pi}\left(\dfrac{q_{o}^2}{2}+\dfrac{k_{B}}{\hbar c}q_{o}T\right)\\
& = \alpha \left(\dfrac{E_{0}}{2k_{B}}+T\right)
\end{align*}
Here  $\alpha=\dfrac{\beta AN_{b} k_{B}q_{0}}{2\pi \hbar c}$ and $ E_{0} = \hbar cq_{0}$. $\;E_{0} \;\text{and} \;q_{0}$ is defined in Figure 6, where $E_{0}$ is qualitatively of the same order as the broadening of the exciton band. $N_{b}$ quantifies the number of phonon branches and are taken to be contributing identically. $A$ is the sample area.

\begin{figure*}[!hbt]
	\centering
%\vspace{-0.1in}
	%\hs{-1in}
	\includegraphics[scale=0.7]{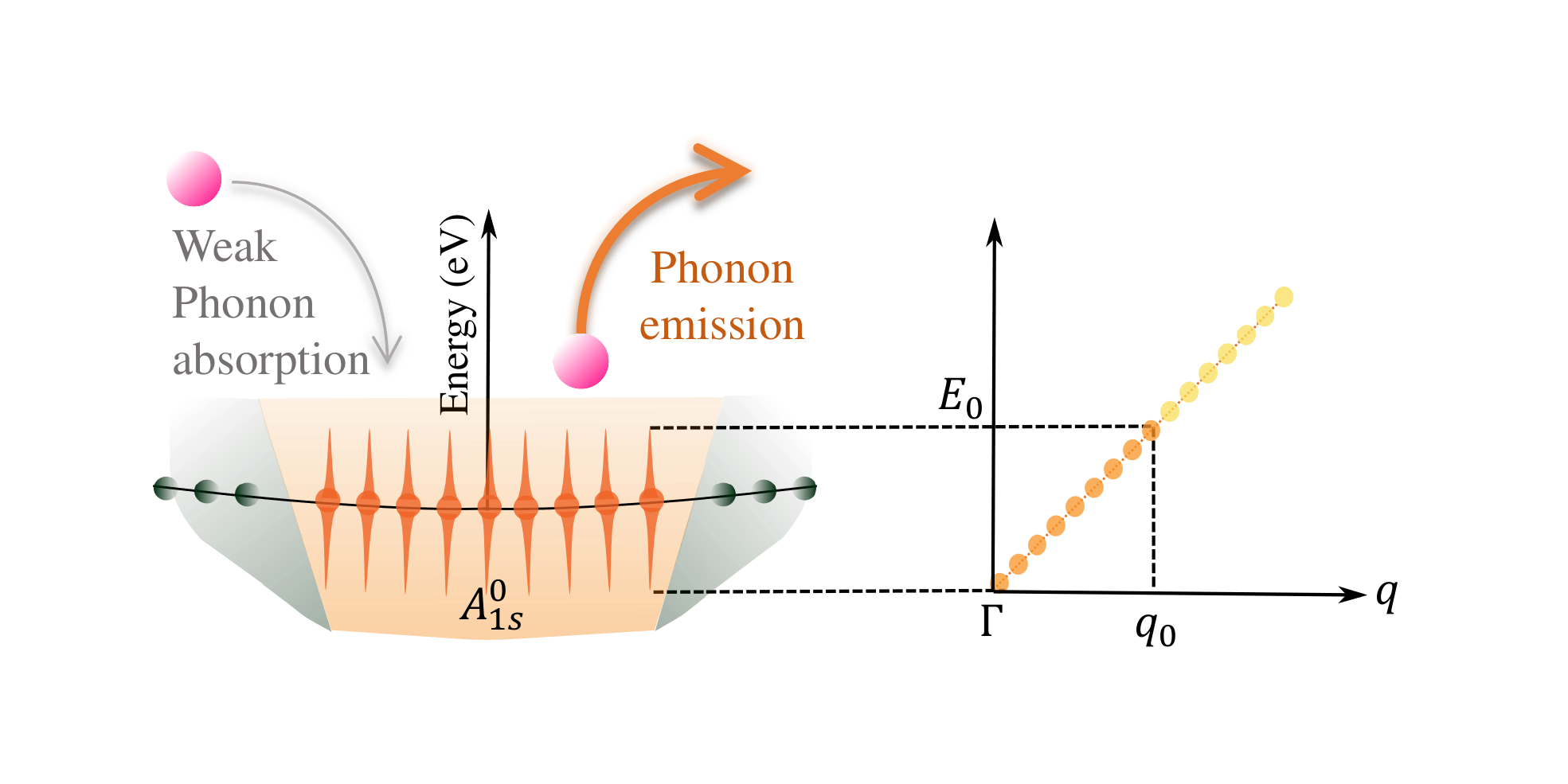}
	%\vspace{-0.1in}
	\caption{\textbf{Non-radiative broadening due to phonon emission as $\textbf{T}\rightarrow\textbf{0}.$} Acoustic modes upto $E_{0}(=\hbar cq_{0})$ contribute to $\Gamma_{NR}$ in the exciton band. Phonon emission is the dominant scattering mechanism (orange arrow) as phonon absorption (grey arrow) is a weak process at small temperatures.}\label{fig:time}
\end{figure*}
\bibliographystylem{ieeetr}
\bibliographym{References}
\renewcommand{\refname}{Bibliography}
\newpage
\subsection*{Supporting Note 1}
	We use an effective low energy Hamiltonian derived from a seven band model using Lowdin partitioning method \cite{kormanyos2015k_si}
	\[
	\begin{pmatrix}
		\varepsilon_v + \alpha k^2& \tau \gamma_3k_{-} + \kappa k_{+}^2 -\tau \dfrac{\eta}{2}k^2k_{+} \\
		\tau \gamma_3k_{+} + \kappa k_{-}^2 -\tau \dfrac{\eta}{2}k^2k_{-}& \varepsilon_c + \beta k^2
	\end{pmatrix}
	\]
	 for calculating the energy dispersion for VB and CB in MoSe\textsubscript{2} around the high symmetry $\textbf{K} (\textbf{K}')$ point in the reciprocal space. It emulates the observed trigonal warping of the isoenergy contours and breaking of electron-hole symmetry around $K$ in DFT calculations\cite{kormanyos2013monolayer_si,wang2016radiative_si}. The basis functions are composed of the predominant Mo $d$ orbitals ($d_{z^2}$ at the CB and $d_{x^2-y^2}, d_{xy}$ at the VB)\cite{xiao2012coupled_si,gong2013magnetoelectric_si} and include relatively smaller yet non-zero contributions from Se $p$ orbitals\cite{liu2015electronic_si,kormanyos2013monolayer_si,kormanyos2015k_si}. We obtain the quasiparticle bandstructure of MoSe\textsubscript{2} and the electronic state in CB(VB) at $\textbf{k}$ as the eigenfunction of the Hamiltonian matrix $\ket*{c,\textbf{k}} (\ket*{v,\textbf{k}}).$
	 \par The exciton state ket $\ket*{{\chi(\textbf{Q})}}$ at centre of mass momentum $\textbf{Q},$ is described as a coherent superposition of electron-hole pairs from all possible band pairs \cite{wu2015exciton_si}
	 \begin{equation}
	 \ket*{\chi(\textbf{Q})} = \sum_{v,c,\textbf{k}} \Lambda(\textbf{Q})\ket*{v,\textbf{k}}\ket*{c,\textbf{k}+\textbf{Q}}.
	 \end{equation}
	 The Bethe-Salpeter (BS) equation is the eigenvalue equation for excitons in the quasiparticle electron-hole basis. The Hamiltonian matrix element includes direct and exchange interaction components given by \cite{wu2015exciton_si,qiu2015nonanalyticity_si}
	 \begin{equation}
	 \mel{vc\textbf{k}\textbf{Q}}{H}{v'c'\textbf{k}'\textbf{Q}} = \delta _{vv'}\delta _{cc'}\delta _{kk'}(\epsilon _{(\textbf{k}+\textbf{Q})c}-\epsilon _{(\textbf{k})v}) - (D-X)_{vv'}^{cc'}(\textbf{k},\textbf{k}',\textbf{Q})
	 \end{equation}
	 \begin{equation}
	 D_{\tau}(\textbf{k},\textbf{k}') = \dfrac{1}{A}V_{\textbf{k}-\textbf{k}'}({}_{\tau}\braket{c,\textbf{k}+\textbf{Q}}{c,\textbf{k}'+\textbf{Q}}_{\tau'}{}_{\tau'}\braket{v,\textbf{k}'}{v,\textbf{k}}_{\tau})
	 \end{equation}
	 \begin{equation}
	  X_{\tau}(\textbf{k},\textbf{k}') = \dfrac{1}{A}V_{\textbf{Q}}({}_{\tau}\braket{c,\textbf{k}+\textbf{Q}}{v,\textbf{k}}_{\tau}{}_{\tau'}\braket{c,\textbf{k}'+\textbf{Q}}{v,\textbf{k}'}_{\tau'})
	 \end{equation}
	 Here $\tau$ corresponds to the valley index, $\tau' = \tau\; (\tau' = -\tau)$ for excitons in the same (time reversal counterpart) valley. We neglect the intervalley and intravalley exchange interactions and obtain exciton band dispersion using direct interaction component for electron-hole pairs within a single valley $\tau$.  $V$ is the intercation potential given by $V_{q}=\dfrac{2\pi e^2}{\epsilon q(1+r_{0}q)},$ $r{_{0}}$ is the screening length depending on the dielectric environment.
	 \par The solution of BS equation is the exciton energy spectrum $E_{s}(\textbf{Q})$ and the probability amplitude $\lambda^{(s)}_{\textbf{Q}}(\textbf{k}),$ where $s\:(n,l)$ stands for exciton band index.
	
	 The momentum matrix element quantifying the electric dipole strength of optical transitions is obtained by \cite{berkelbach2015bright_si}
	 \begin{equation}
	 \textbf{P}_{vc,\textbf{Q}}(\textbf{k})=\mel*{v,\textbf{k}}{\textbf{p}}{c,\textbf{k}+\textbf{Q}} = \dfrac{m}{\hbar}(\epsilon_c(\textbf{k}+\textbf{Q})-\epsilon_v(\textbf{k}))\mel{v,\textbf{k}}{\pdv*{\textbf{k}}}{c,\textbf{k}+\textbf{Q}}
	 \end{equation}

	 \par The quantity $\chi_{ex}(\textbf{Q})$ can be calculated using the following expression \cite{palummo2015exciton_si,wang2016radiative_si}
	 \begin{equation}
	 \chi_{ex}(\textbf{Q}) = \int \dfrac{d^{2}\textbf{k}}{(2\pi)^2}\textbf{P}_{vc,\textbf{Q}}(\textbf{k}).\:\hat{x}\: \lambda^{(s)}_{\textbf{Q}}(\textbf{k})
	 \end{equation}
	 \par
	 \par We use the k.p Hamiltonian parameters derived using $d$ orbital contribution from ref \cite{xiao2012coupled_si} for MoSe\textsubscript{2}. The additional variables due to Lowdin approximation are the same as that for MoS\textsubscript{2} \cite{kormanyos2013monolayer_si}. We verify an insignificant change in the final results due to deviation in the Lowdin parameters. The validity of this approximation also lies in relative atomic orbital contribution of chalcogen $p$ and Mo $d$ at the $K(K')$ point, which is almost similar for MoS\textsubscript{2} and MoSe\textsubscript{2} \cite{kosmider2013large_si}.  The error bar quantifies the deviation in the final results due to (a) $\pm 5\%$ variation Lowdin parameters and (b) effect of $\textbf{k}$-space sampling on the solutions of BSE \cite{qiu2013optical_si}.  Note that $E_{1s}(\textbf{0})$  was maintained at the experimental $1s$ peak position in all the cases by varying $r_{0}, \epsilon$ in the interaction potential.
	
	\pagebreak
	\subsection*{Supplementary Note 2}
	\begin{figure*}[!hbt]
		\centering
		%	\vspace{-0.1in}
		%\hs{-1in}
		\includegraphics[scale=0.7] {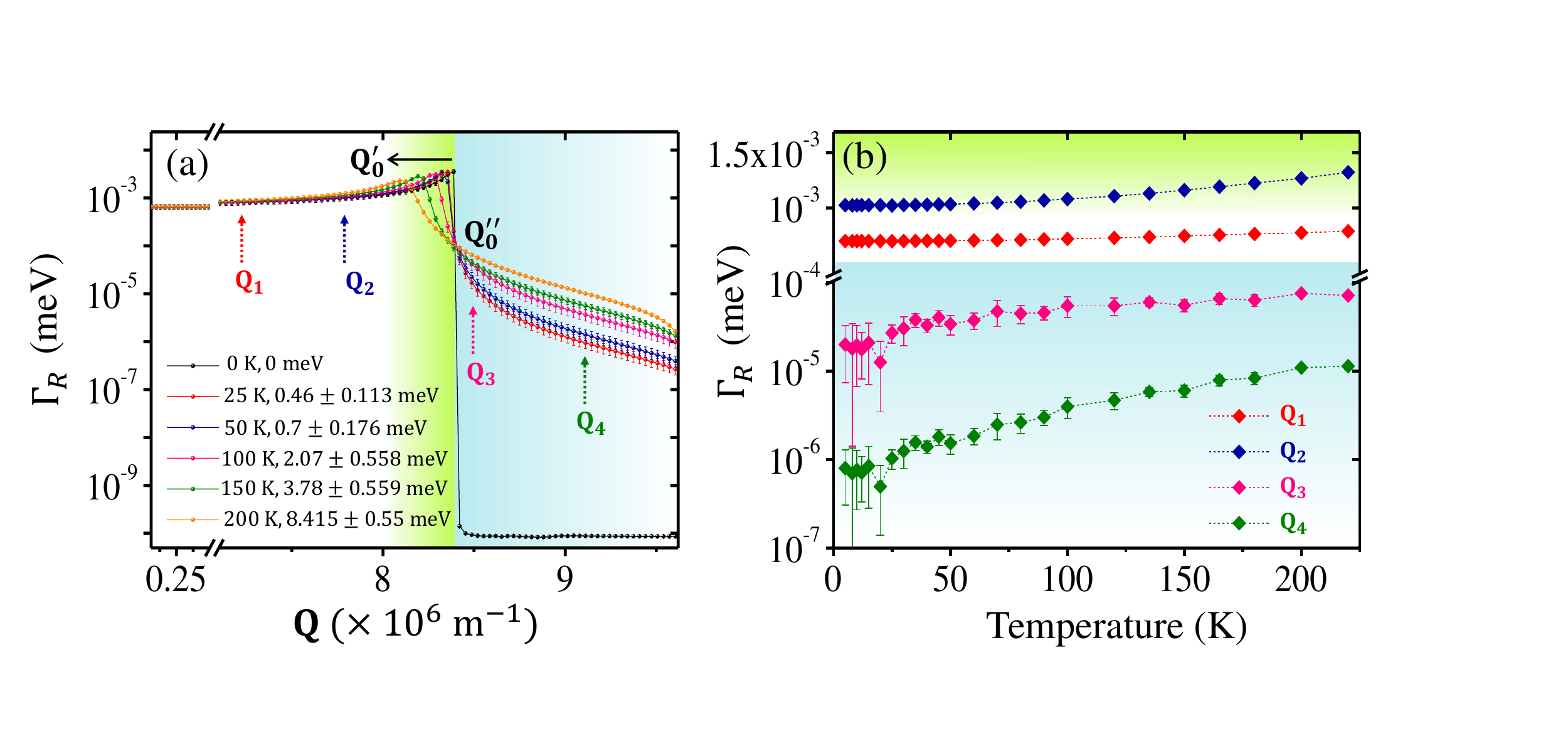}
		%\vspace{-0.1in}
		\caption{\textbf{Effect of temperature on radiative broadening }(a) $\Gamma_{R}$ vs $\textbf{Q}$ as the temperature changes (see main text). (b) The change in $\Gamma_{R}$ with temperature for the pointed $\textbf{Q}$ points in (a)}\label{fig:time}
	\end{figure*}

The relative increment in  $\Gamma_{R}$ is approximately $7.53\: \%, 27.32\: \%$ for states $\textbf{Q}_{\textbf{1}},\textbf{Q}_{\textbf{2}}  \;(\textbf{Q}_{\textbf{1}} < \textbf{Q}_{\textbf{2}} < \textbf{Q}_\textbf{0}^{'})$ respectively on changing the sample temperature from 5 K to 220 K. The enhanced increment for $\textbf{Q}_{\textbf{2}}$ as compared to $\textbf{Q}_{\textbf{1}}$ is due to the left-shift in $\textbf{Q}_{\textbf{0}}^{'}$ with temperature. On the other side of the light cone, as a result of increase in exciton contribution in the light cone due to $\Gamma_{NR},$ the percentage change increases to $2\times 10^2\: \%, 1.33\times 10^3\: \%$ for  $\textbf{Q}_{\textbf{3}}, \textbf{Q}_{\textbf{4}}  \;( \textbf{Q}_{\textbf{4}} > \textbf{Q}_{\textbf{3}} > \textbf{Q}_\textbf{0}^{'})$ respectively. The relative change is large for $\textbf{Q}$ states near to $\textbf{Q}_{\textbf{0}}^{'}$ and outside of it a little, whereas the points lying inside are mostly unaffected as their exciton contribution isn't changed much by the lower bound energy line $\hbar cQ.$
%	\pagebreak

  \bibliography{References_Supp_info}
 \bibliographystyle{ieeetr}
\end{document}